# Improvements in Tropical Cyclone Representation in JRA-3Q over ERA5

**Yushan Han[1], Paul A. Ullrich[1,2], Colin Zarzycki[3], Stella Bourdin[4]**

[1]Department of Land, Air and Water Resources, University of California, Davis, California, USA

[2]Division of Physical and Life Sciences, Lawrence Livermore National Laboratory, Livermore, CA, USA

[3]Department of Meteorology and Atmospheric Science, Pennsylvania State University, State College, PA, USA

[4]LMD/IPSL, École Polytechnique, Institut Polytechnique de Paris, ENS, PSL Research University, Sorbonne Université, CNRS, Palaiseau, France

Corresponding author: Yushan Han (yshhan@ucdavis.edu)

**Key Points:**

- JRA-3Q represents TCs more accurately than ERA5, with higher hit rates, fewer false alarms, and more stable long-term performance
- JRA-3Q's improved TC representation likely results from both reduced model biases and a more effective TC bogus scheme
- TC representation biases arising from tracking algorithms, reanalyses, and IBTrACS are systematically decomposed

## Abstract

Reanalysis products are widely used for long-term tropical cyclone (TC) studies, yet the fidelity of their TC representation is not fully established. We evaluate TC representation in the recently released JRA-3Q reanalysis against the higher-resolution ERA5 using the SyCLoPS tracker. JRA-3Q achieves substantially higher detection skill, with a hit rate of 90% and a false alarm rate of 6% during 1979–2024. Unlike ERA5, which exhibits markedly degraded detection skill during the pre-satellite era, JRA-3Q maintains stable performance over time. We analyze false alarms and missed TCs in the North Indian basin, TC vortex evolution following TC bogussing, and the spatial variance of column-integrated moist static energy, showing that JRA-3Q's improvements stem from both reduced model biases and a more effective and consistently applied TC bogus scheme. A decomposition of TC representation errors into contributions from tracking algorithm, reanalysis, and best-track uncertainty helps explain the sources of these biases.

## Plain Language Summary

Reanalyses combine observations with numerical weather models to reconstruct past atmospheric conditions and are widely used to study long-term tropical cyclone (TC) variability. However, it is still unclear how well reanalysis products actually represent TCs. We compare TC representation in two modern reanalysis products: the recently released JRA-3Q reanalysis and the widely used ERA5 reanalysis. Using an objective TC tracker, we detect more observed TCs in JRA-3Q and find fewer false detections when compared with ERA5. JRA-3Q also maintains more consistent performance before and after the satellite era. We find that these improvements result from both reduced model biases and a more effective TC bogus scheme, which inserts synthetic TC information into the data assimilation system when observations are sparse. We also identify the main sources of errors in TC representation, including limitations of tracking algorithms, the reanalyses themselves, and uncertainties in historical TC best-track records. Our results provide insights into selecting and utilizing reanalysis products for future TC research.

## 1. Introduction

Tropical cyclones (TCs) are a central focus of climate and Earth system modeling studies because they are among the most devastating weather-related disasters during 2000-2019 (CRED & UNDRR, 2020). The International Best Track Archive for Climate Stewardship (IBTrACS; Gahtan et al., 2024; Knapp et al., 2010) is a global collection of observed TC data, but it contains uncertainties, particularly in earlier decades when observing systems and operational practices were more limited (Kossin et al., 2013; Vecchi & Knutson, 2011). Consequently, previous studies have highlighted the value of objectively identifying TCs in reanalysis products to augment confidence in best-track datasets by uncovering historically undetected TCs (Chand et al., 2022; Hodges et al., 2017; Truchelut et al., 2013). Recent advances in reanalysis have enabled a wide range of climatological and physical studies of TCs in the context of their large-scale environment (e.g., Bhatia et al., 2022; Carstens et al., 2024; Chavas et al., 2025; Sobel et al., 2021). However, because reanalyses are produced with general circulation models only partially constrained by observations, understanding their limitations in representing TCs is important.

The fifth-generation European Centre for Medium-Range Weather Forecasts reanalysis (ERA5; Hersbach et al., 2020; Soci et al., 2024) is the most widely used reanalysis product for TC studies. Its relatively high horizontal and vertical resolution, state-of-the-art model physics and data assimilation, and extended temporal coverage facilitate advanced TC representations (e.g., Bhatia et al., 2022; Zhang & Chan, 2023; Dulac et al., 2024; Zarzycki et al., 2021). ERA5 is also widely used to train modern AI weather and climate models (e.g., Alet et al., 2026; Kochkov et al., 2024; Watt-Meyer et al., 2025). Compared with previous generations of reanalyses, ERA5 reproduces more TC and monsoonal low-pressure system (LPS) tracks (Vishnu et al., 2020; Zarzycki et al., 2021) and better represents TC size and structure (Bian et al., 2021; Carstens et al., 2024; Malakar et al., 2020). However, ERA5's shortcomings are also well documented, particularly its underestimation of TC intensity (Dulac et al., 2024; Xu et al., 2024) and biases in environmental fields (Slocum et al., 2022). Dirkes et al. (2023) further showed that ERA5 does not necessarily perform better for TC moist static energy (MSE) variance budget profiles. Previous studies have also documented overestimated TC frequency in the Indian Ocean and underestimated frequency in the North Atlantic (Bourdin et al., 2022; Han & Ullrich, 2026b; M. Roberts et al., 2020), although it is unclear if the error can be fully attributed to reanalysis.

The recently introduced Japanese Reanalysis for Three Quarters of a Century (JRA-3Q; Kosaka et al., 2024) provides new opportunities for long-term TC analysis at a higher horizontal resolution than its predecessor, Japanese 55-year Reanalysis (JRA-55) (0.375° versus 0.563°), although it remains coarser than ERA5 (0.25°). JRA-55 performs competitively in representing TCs (Hodges et al., 2017; Murakami, 2014; M. Roberts et al., 2020; Zarzycki et al., 2021) but suffers from an artificial long-term decline in TC intensity caused by assimilating wind-profile retrievals around TCs Kobayashi et al., 2015). JRA-3Q addresses this bias with a new TC bogus scheme based on IBTrACS, yielding a higher and more stable TC detection rate (i.e., hit rate) throughout the reanalysis period (Kosaka et al., 2024). ERA5 also assimilates IBTrACS bogus observations during its back-extension period (1940–1978), but with a substantially different implementation.

Using the physics-based, machine learning-aided LPS tracker SyCLoPS (Han & Ullrich, 2025) for TC detection in reanalysis products, we find that JRA-3Q not only represents TC intensity more accurately than ERA5 and achieves a substantially higher Hit Rate (HR)—likely owing to its TC bogus scheme—but also markedly reduces the false alarm rate (FAR) across all basins—an improvement that cannot be explained by TC bogussing alone. JRA-3Q also substantially outperforms ERA5 during the pre-satellite era when both reanalyses employ TC bogus observations. These initial results motivate a detailed comparison of TC representation in the two reanalyses in Section 3, quantified by the HR and FAR, and an investigation of the factors contributing to JRA-3Q's improved performance. In Section 4, we further decompose the sources of error contributing to TC representation biases.

## 2. Data and Methods

This study uses three reanalysis products: ERA5 from 1940 to 2024 on a 0.25º × 0.25º rectilinear grid; JRA-3Q from 1948 to 2024 (0.375º × 0.375º). The overlapping analysis period is 1948–2024. The six major TC basins—North Atlantic (NA), eastern Pacific (EP), western Pacific (WP), North Indian (NI), South Indian (SI), and South Pacific (SP)—are analyzed separately for each reanalysis. Basin boundaries are shown in Figure S1. TC observations are obtained from IBTrACS. We remove tracks that never reach a 10-m wind speed of 17.5 m/s. To minimize interagency heterogeneities and maximize coverage across eras and basins, track filtering uses only wind speeds from U.S. sources. Different data sources yield similar results but slightly higher FAR (see Table S1).

The System for Classification of Low-pressure Systems (SyCLoPS; Han & Ullrich, 2025) is the primary TC tracker used for all three reanalyses at 6-hourly intervals because of its superior detection skill and comprehensive LPS classification ability across system types and intensities. In this study, this framework facilitates diagnosis of the LPS characteristics associated with missed TC tracks and TC precursors in reanalyses. The conventional Zarzycki and Ullrich (ZU; Zarzycki & Ullrich, 2017) tracker is included for comparison, while noting that ZU was originally tuned using an older generation of reanalysis products. Both trackers are built on TempestExtremes (Ullrich et al., 2021) and select TCs primarily based on warm core criteria and mean sea-level pressure (MSLP) gradient, with the ZU algorithm modified by removing its surface elevation constraint to ensure consistency with SyCLoPS. Details of these two tracking algorithms can be found in the appendices of Han & Ullrich (2026).

The HR and miss rate (MR) are defined as the proportion of IBTrACS tracks detected and missed by the tracking algorithm, respectively (i.e., hits and misses). A hit occurs when a detected TC track contains a node within 2° great-circle distance of an IBTrACS record at the same timestamp. TC detection is not sensitive to the choice of this matching distance (Bourdin et al., 2022). FAR is defined as the proportion of tracks identified by the tracking algorithm that are not found in IBTrACS; that is, tracks with no detected TC nodes within 2° of any IBTrACS record. TC detection skill is summarized by HR minus FAR (HRMFAR).

Previous studies using different trackers have shown lower HRs in the NA compared to the global mean when evaluated against IBTrACS observations (Bourdin et al., 2022; Han & Ullrich, 2026; Ye et al., 2026). Han & Ullrich (2026) found that many of these missing tracks are detected as subtropical storms in SyCLoPS. To address this limitation, we update SyCLoPS to recover as many missing tracks as possible while avoiding introducing FAs. Specifically, we do so by identifying a class of "hybrid TCs" that fail SyCLoPS's tropical 100-hPa humidity criterion but satisfy the remaining TC requirements. These systems occur predominantly in the NA and are counted as TC tracks in the NA. Details on hybrid TC detection are provided in Text S1 and Figure S2 in Supporting Information S1.

The column-integrated Moist static energy (MSE; $\hat{h}$) analyzed in this study is calculated as:

$$\hat{h} = \int_{p_t}^{p_b} (C_p T + gZ + L_v q)\ dp \quad (1)$$

where $p_b$ and $p_t$ are the bottom and top pressure levels, here chosen as 925 and 1 hPa, as suggested by Wing et al. (2019). $q$ is the water vapor mixing ratio, and the other terms have their conventional definitions. The spatial variance of $\hat{h}$ is defined as the square of $\hat{h}'$. The prime indicates the deviation from the average of a 10º box centered at TC locations.

**3. Results**

We first evaluate how well JRA-3Q, ERA5, and JRA-55 represent TC intensity using frequency distributions and wind–pressure relationships (Figure S3). Overall, JRA-3Q most closely reproduces the observed intensity distribution, while ERA5 exhibits a larger underestimate of near-surface wind speeds over the same range of MSLP. All three reanalyses reproduce the observed peak in minimum MSLP near 990–1000 hPa but underestimate the low-MSLP tail associated with the strongest TCs.

The first and third columns of Figure 1 show the nine-year moving average time series of detection skill (HRMFAR) of SyCLoPS and ZU in each reanalysis. Overall, SyCLoPS demonstrates improved TC detection skill over ZU across most basins in both reanalyses, primarily through reducing FAs, especially in the SP and EP, where subtropical/extratropical cyclones and weak tropical systems have been reduced. SyCLoPS also removes dozens of FAs in the South Atlantic basin that are operationally considered subtropical cyclones (Evans and Braun, 2012) (not shown). Globally, JRA-3Q attains an HR of 90% and a FAR of 6% during the satellite era (1979–2024), representing, to our knowledge, the best TC detection skill scores reported for a global reanalysis. The corresponding values for ERA5 are 78% and 11%. In the NA, SyCLoPS further improves both HR and FAR by including hybrid TC tracks. The benefit is substantial for JRA-3Q but only modest for ERA5.

As shown in Figure 1, JRA-3Q outperforms ERA5 across all basins and both tracking algorithms, owing to lower FAR and more hits. In the well-observed NA, EP, and WP basins, JRA-3Q maintains low FAR, with HRMFAR approaching or exceeding 90% (Figs. 1a, c, e). The NI basin exhibits the largest advantage in FAR for JRA-3Q, yet its peak HRMFAR remains lower than in the other basins (Fig. 1g). JRA-3Q's detection skill is also relatively stable throughout the record in the Northern Hemisphere. In contrast, ERA5 improves markedly from the pre-satellite (1948–1978) to satellite (1979–2024) era across basins, with HR increases continuously in some cases as the observing system advances. This contrast is particularly pronounced in the EP (Fig. 1c), where ERA5's low pre-satellite HRMFAR reflects few hits and a surge in FAs (Fig. 1d). Because both reanalyses employ TC bogussing during this period, this difference suggests that the ERA5 scheme is less effective than JRA-3Q's in compensating for sparse conventional observations. JRA-55 shows the opposite temporal evolution, with initially high detection skill that gradually declines, consistent with Kosaka et al. (2024) (Fig. S4).

The observed TC numbers clearly increase with time in several basins, which is captured by both reanalyses. However, a substantially lower number of observed TC in the pre-satellite era may reflect false negatives in IBTrACS. For example, in the SI and SP, both reanalyses have substantially lower HRMFAR (Figs. 1i, k) and a greater number of FAs (Figs. 1j, l) during the pre-satellite era, suggesting that part of the apparent degradation in HRMFAR comes from limitations in historical IBTrACS observations. Some of the FAs identified in both reanalyses may therefore represent TCs that were missed by operational agencies.

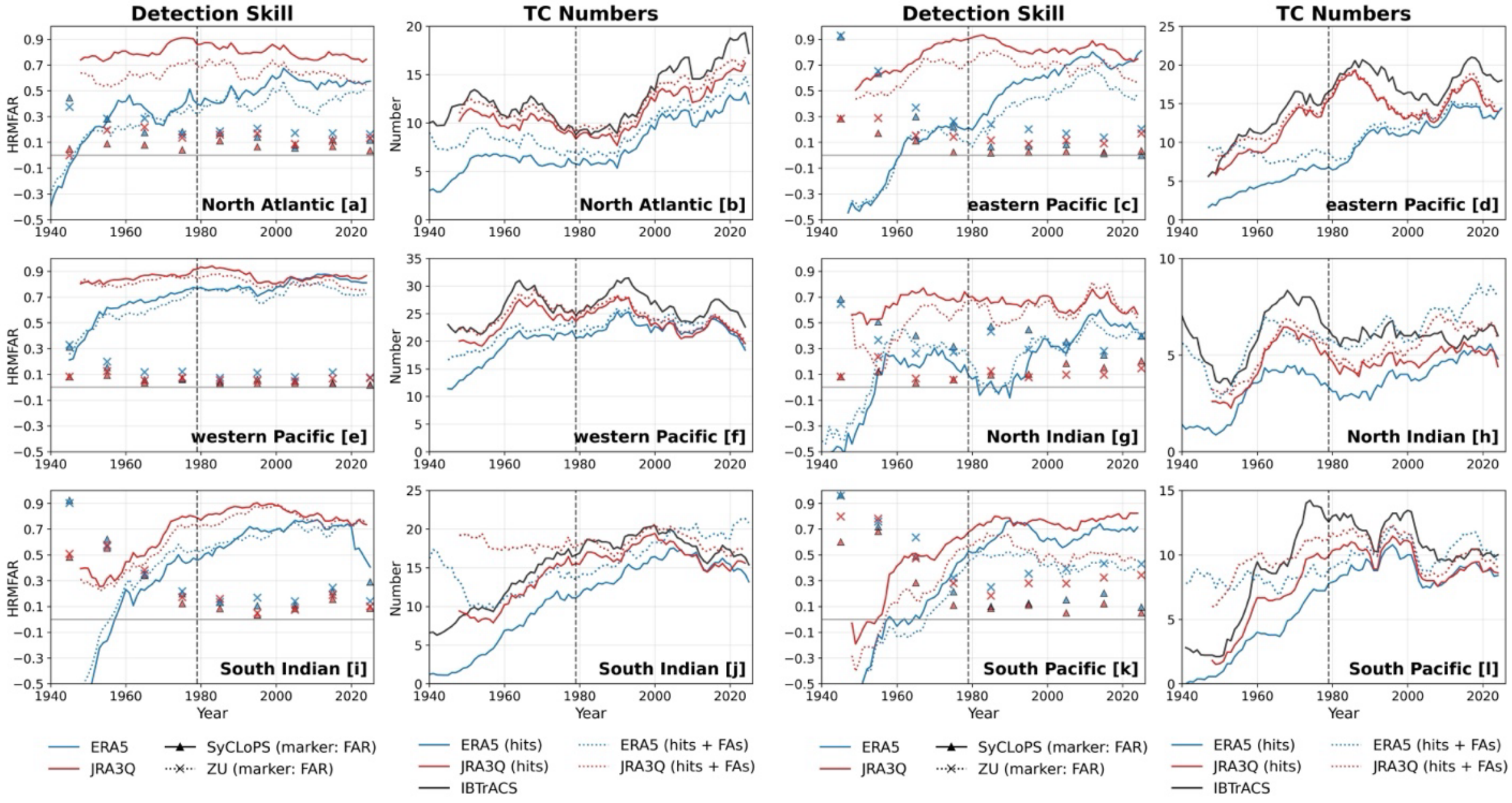

**Figure 1.** Nine-year moving average time series of TC numbers (solid lines for hits and dotted lines for hits plus false alarms) and TC detection skill HRMFAR (solid lines for SyCLoPS and dotted lines for ZU) with decadal-mean FAR (markers) for ERA5 (blue) and JRA-3Q (red). The vertical dotted line separates the pre-satellite and satellite era.

Global differences in true-positive TC node density and false alarm (FA) density between JRA-3Q and ERA5 during the satellite era are shown in Figures 2a and 2b. JRA-3Q produces substantially more hits throughout the tropics and fewer FAs across most basins, with the largest FA reductions over the Indo-Gangetic Plain and northwestern Australia. Its advantage is particularly evident in the EP, while the subtropical NA also shows substantially more hits. Conversely, JRA-3Q produces fewer hits in the midlatitude WP and subtropical SP, although its lower FA counts partly offset these differences, potentially reflecting uncertainty in the SyCLoPS threshold defining extratropical transition.

Interestingly, in the NI, ERA5 produces slightly more hits over the coastal Bay of Bengal but also many more FAs. The NI contains many monsoonal systems that can intensify near and over land; while hybrid cyclones are uncommon, it is thus an ideal basin for examining TC representation over tropical land and ocean in the monsoon environment. We therefore examine reanalysis-unique hits and land FAs within the gray rectangle in Figures 2a and 2b. ERA5 produces substantially more unique land FAs than JRA-3Q (Figs. 2c, d). Although JRA-3Q also produces some land FAs, they are less concentrated west of 80°E and over the western Ganges Delta, where monsoon depressions frequently propagate (Vishnu et al., 2020). These ERA5 FAs span a broad intensity range (980–1005 hPa) and exhibit mature, asymmetric cyclone structures, with strong ascent and elevated near-surface equivalent potential temperature near the center (Fig. 2e). Their asymmetric, westward-tilting circulation resembles that described by Hunt et al. (2016), potentially reflecting Himalayan influences that modify the circulation and suppress convection on the eastern flank. At the same locations and times, JRA-3Q represents these systems as substantially weaker and less organized (Fig. 2f). Conversely, JRA-3Q-unique FAs are generally weaker and more symmetric (Fig. 2g). These results suggest that ERA5 is more prone to producing overly strong circulations over the NI than JRA-3Q, despite assimilating broadly similar observations.

IBTrACS tracks uniquely reproduced in ERA5 are considerably fewer than those uniquely reproduced in JRA-3Q, and the two reanalyses capture distinct subsets of tracks (Figs. 2c, d). ERA5 unique hits are dominated by northwestward-moving monsoon depressions from the northern Bay of Bengal, whereas JRA-3Q unique hits originate farther south, reach higher intensities, and include substantially more Arabian Sea TCs. Their vertical structures further differ: ERA5 unique hits closely resemble its FAs but with slightly stronger circulation (Fig. 2i), while these systems in JRA-3Q have slightly weaker wind speeds and a similar asymmetric structure (Fig. 2j). In contrast, JRA-3Q unique hits exhibit a mature, nearly symmetric vortex with near-surface winds exceeding 18 m s⁻¹, while corresponding systems are much weaker and less organized in ERA5 (Figs. 2k, l).

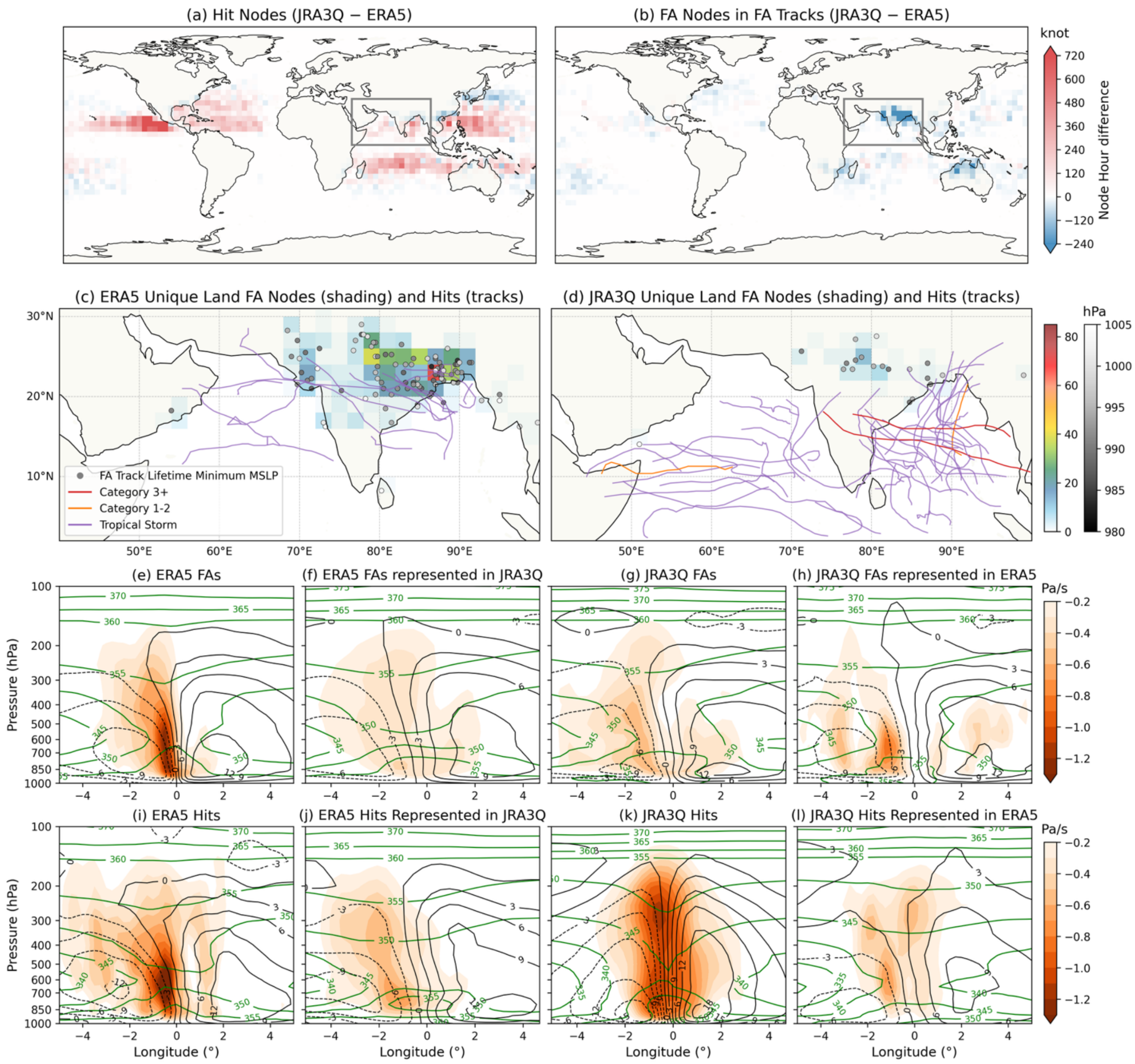

**Figure 2.** 4° × 4° density (represented by TC node hours, one node is 6 hours) of (a) hits and (b) FAs between JRA-3Q and ERA5. Unique hits (IBTrACS tracks) and Land FAs (2° × 2° shading) over the NI during the satellite era in (c) ERA5 and (d) JRA-3Q. The last two rows show 10° west–east vertical composites centered on the times and locations of the observed lifetime peak wind speed of hit tracks (e–h) and lifetime minimum MSLP of FA tracks (i–l), with shading indicating vertical velocity, black contours the meridional wind (m s⁻¹), and green contours equivalent potential temperature (K).

Reanalyses employing TC vortex bogussing usually outperform those without it when comparing TC intensities (Schenkel & Hart, 2012; Zick & Matyas, 2015). This suggests that the TC bogus scheme in JRA-3Q contributes to its superior reproduction of observed TCs. Although ERA5 also assimilates TC bogus data derived from IBTrACS MSLP observations in the pre-satellite period, the two systems differ substantially in implementation. ERA5 assimilates each IBTrACS MSLP record as a single-point observation, whereas JRA-3Q constructs a synthetic three-dimensional TC vortex using the reported MSLP, an empirically estimated gale-force wind radius, gradient wind balance, and prescribed wind and pressure profiles. Observations sampled from the vortex center and surrounding locations are then assimilated as TC bogus data (Kosaka et al., 2024).

Figures 3a–h compare the impacts of TC bogussing in the WP during 1951–1978, when both reanalyses assimilate TC bogus observations and relatively consistent MSLP observations are available. We define a "hard bogus" as tracks lacking any detectable LPS in SyCLoPS within 12 h prior to the first MSLP record in the corresponding IBTrACS track (i.e., cases where bogussing is obvious at the first observation); all other tracks are classified as "soft-bogus" tracks. In JRA-3Q hard-bogus tracks, neither a closed MSLP contour nor a coherent elevated 850-hPa relative vorticity region exists before bogussing, yet within six hours the vortex develops dramatically into a well-organized cyclone with a distinct inner core. In contrast, ERA5 hard-bogus cases show only modest vortex intensification, with slightly stronger low-level vorticity and limited MSLP deepening. The effect is even weaker for ERA5 soft-bogus cases, whose vortices are already well developed before the first IBTrACS record. Although JRA-3Q soft-bogus cases also contain preexisting weak vortices and exhibit only slight average deepening of minimum MSLP at +0 h, they undergo substantially greater structural development following bogussing than their ERA5 counterparts.

Figures 3i–p show the spatial variance of column-integrated MSE ($\hat{h}'$) for 100 randomly selected weaker (985–995 hPa) and stronger (935–945 hPa) Northern Hemisphere TCs across the two reanalyses and eras. Under weak temperature gradients, $\hat{h}'$ reflects moisture variability associated with organized convection (Carstens & Wing, 2022). ERA5 consistently has lower inner-core $\hat{h}'$ than JRA-3Q, suggesting weaker convective and diabatic activity, although differences may also reflect MSE advection, assimilation tendencies, and other unresolved processes (Dirkes et al., 2023). ERA5 also exhibits substantially higher $\hat{h}'$ in the satellite era, suggesting that improvements in its representation of TC structure may contribute to its increasingly higher HRMFAR over time. In contrast, JRA-3Q remains relatively consistent over the two periods, suggesting reduced sensitivity to observing-system changes likely associated with its consistent use of TC bogussing.

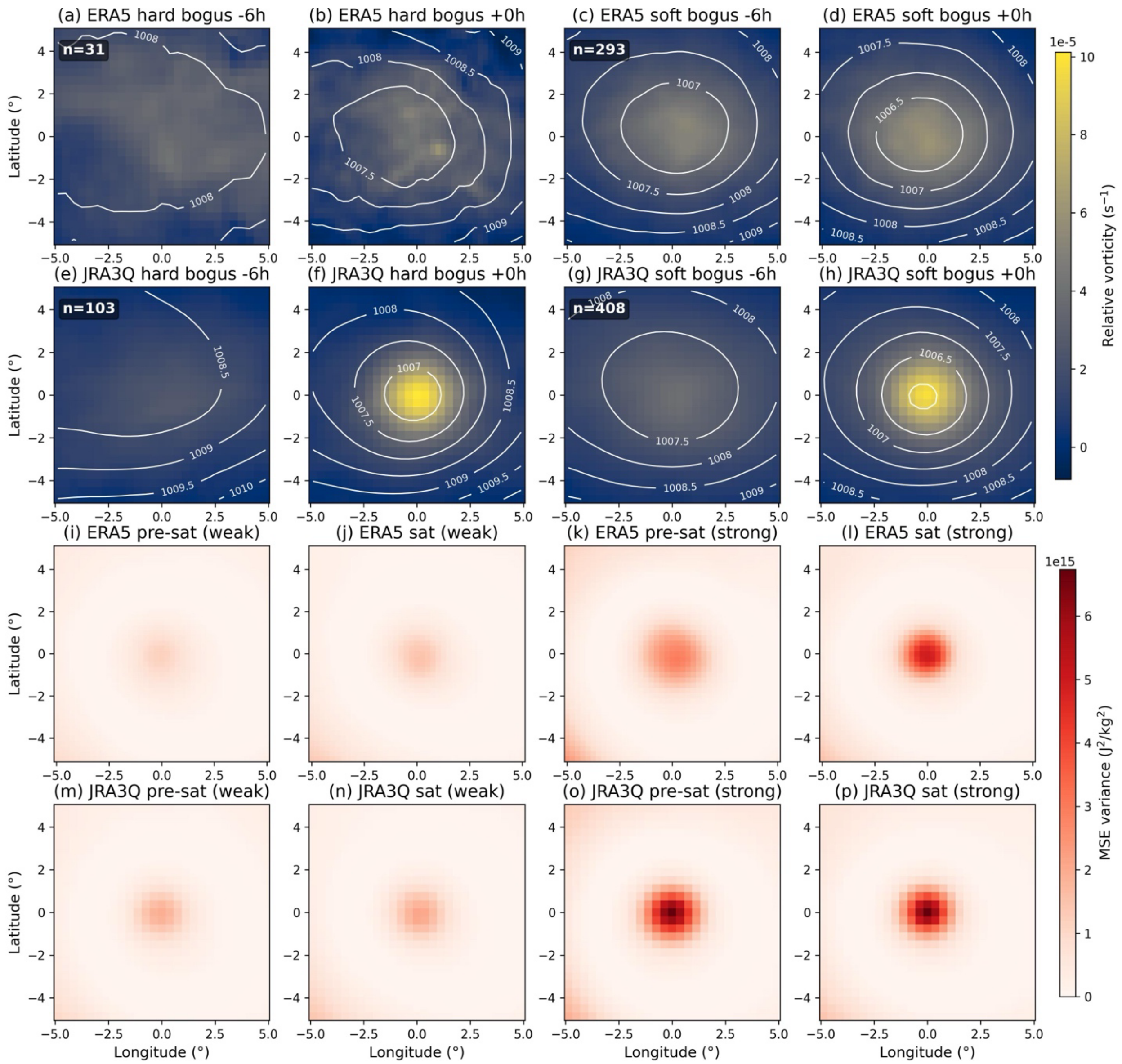

**Figure 3**. (a–h) Composite of MSLP contours (white) and 850 hPa relative vorticity (shading) with sample size denoted for each group of cases. (i–p) Composite of $\hat{h}'$ for 100 randomly selected TCs. See the text for details.

Ultimately, misses and FAs in reanalysis arise from three broad sources: (1) tracking algorithm biases, (2) reanalysis deficiencies, including model biases, TC bogussing, and assimilation processes, and (3) uncertainties in the IBTrACS best-track dataset. Figure 4 qualitatively decomposes these sources for JRA-3Q and ERA5 by leveraging the SyCLoPS LPS catalog and IBTrACS without filtering non-TCs. Although the individual contributions cannot be fully isolated, this decomposition provides insight into their relative importance.

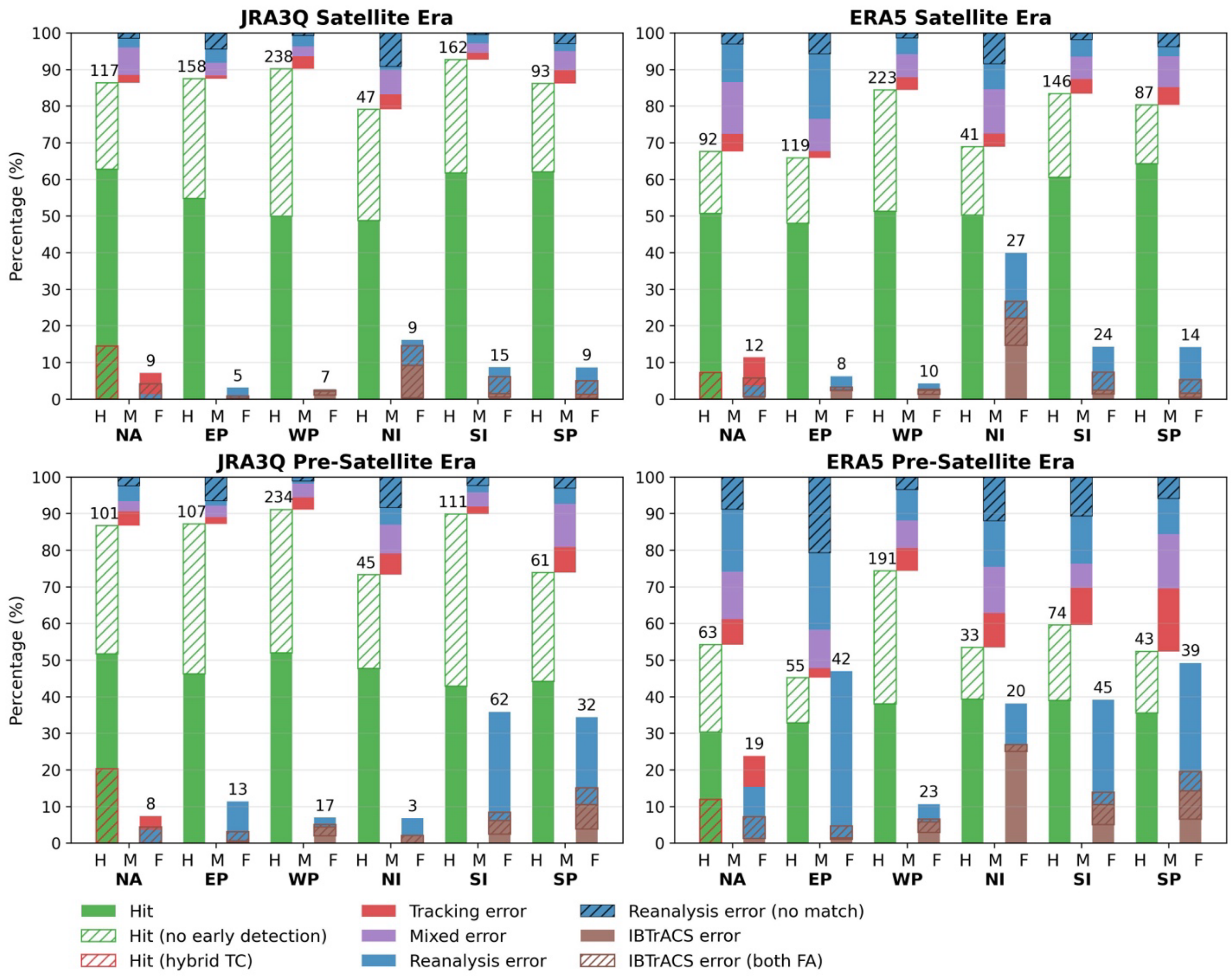

**Figure 4**. Decomposition of error sources and uncertainties in the hit rate (H), missing rate (M), and false alarms (F) in ERA5 and JRA3Q over the satellite and pre-satellite eras across the six basins. Numbers over the bars are decadal-mean TC counts. See the text for more details.

Missed TCs attributed to tracking errors are those classified by SyCLoPS as non-tropical systems or tropical depressions for at least one day, which are most sensitive to classification thresholds. Cases failing only the one-day TC duration criterion are classified as mixed errors because the minimum duration required to define TC formation is inherently subjective. Remaining misses are attributed to reanalysis deficiencies because they are classified as weak tropical systems without a detectable upper-level warm core. Black hatching indicates cases without a matchable SyCLoPS LPS. Because FAR is nearly zero in the well-observed NA, EP, and WP, tracking-related FAR is assumed negligible except for hybrid TC-related FAs in the NA. FAs matched to IBTrACS tracks that fail the 34-kt threshold are classified as potential IBTrACS errors, particularly when reproduced by both reanalyses (brown hatching) or occurring in the pre-satellite era. Green hatching in the HR column indicates cases lacking detectable precursors in both SyCLoPS and a low-level relative-vorticity tracker (Hodges et al., 2017; Bourdin, 2026) before the first IBTrACS observation, suggesting that these TCs may have emerged from the TC bogus scheme.

Figure 4 shows that JRA-3Q generally has smaller "pure" reanalysis errors in both misses and FAs than ERA5, consistent with its lower MR and FAR across most basins. The improvements in MR and FAR can be broadly linked to differences in JRA-3Q's TC bogus scheme and underlying model, respectively. JRA-3Q's HR also remains relatively stable between the two eras, whereas ERA5 degrades substantially in the pre-satellite era, particularly in the EP, where both MR and FAR increase. This degradation likely reflects sparser observations and weaker TC bogussing influence, allowing ERA5 model biases to become more apparent. However, JRA-3Q's larger fraction of hits without detectable pre-TC disturbances suggests its greater reliance on TC bogussing for vortex initiation. Neither reanalysis recovers the artificially low observed TC frequency in the EP during the pre-satellite era (~12 annually versus ~18 in the satellite era), even with FAs included.

Despite JRA-3Q's overall higher HR, its unusually low FAR in the NI, particularly during the pre-satellite era (~0.3 $yr^{-1}$), warrants caution given greater IBTrACS uncertainty (Fig. 4). It is possible that some ERA5 FAs represent undocumented TCs, while JRA-3Q misses TCs that are also absent from IBTrACS, suggesting that ERA5 may overestimate whereas JRA-3Q may underestimate the intensity of some NI systems. Figure 2g supports this interpretation, showing several inland-moving monsoonal systems captured by ERA5 but missed by JRA-3Q. This is also consistent with the larger number of JRA-3Q misses with no detectable match, and the better NI TC intensity and structural representation reported for ERA5 by Malakar et al. (2020). These inland systems are also subject to best-track uncertainty because TC-like structures may persist even when land friction prevents sustained surface winds from reaching operational thresholds.

## 4. Conclusions

This study demonstrates substantial improvements in TC representation in JRA-3Q relative to ERA5. JRA-3Q achieves higher detection skill across all basins and maintains relatively stable performance in Northern Hemisphere basins with better observational coverage. In contrast, ERA5 shows an anomalous increase in HRMFAR across both the pre-satellite and satellite eras, warranting caution when using ERA5 for climatological TC analysis. In the Southern Hemisphere, both reanalyses show substantially degraded performance, likely reflecting a combination of missing TC tracks in IBTrACS and inherent limitations in representing TCs under sparse observations. The advantages of JRA-3Q over ERA5 are further revealed by a global increase in hits and reduction in FAs, particularly over coastal monsoon regions, as well as greater and more consistent spatial variance in inner-core column-integrated MSE across intensity categories and observational eras, suggesting a more stable TC representation.

Our results also highlight potential limitations of TC bogussing in JRA-3Q. Mature JRA-3Q vortices develop unphysically from weak or undetectable precursors following bogussing, suggesting overfitting to pseudo-observations derived from IBTrACS. JRA-3Q's unusually low FAR in the NI and larger fraction of tracks without detectable pre-TC disturbances further reveal potential model limitations where IBTrACS records are incomplete. In ERA5, the weaker influence of bogus observations may reflect the reduced assimilation weight applied relative to the preliminary ERA5 extension to avoid unrealistically strong TCs during the back-extension period (Bell et al., 2021; Soci et al., 2024). Overall, these results underscore the challenge of balancing TC bogussing against deficiencies in the underlying model and other components of the reanalysis system.

We acknowledge that further diagnosis and decomposition of these TC representation deficiencies require process-oriented analysis of the reanalyses and, more importantly, their underlying general circulation models. Our preliminary analysis shows that ECMWF-IFS-HR (C. Roberts et al., 2017), a HighResMIP model sharing the underlying model of ERA5, exhibits similar biases in its uncoupled historical run over a 27-year period, including an underrepresentation of EP and NA TCs and an overrepresentation of inland NI TCs (Figure S5). Similar behavior is found in some, but not all, HighResMIP models. ECMWF-IFS-HR also overproduces TCs along the tropical West African coast and adjacent inland regions, although whether this reflects the same mechanism as the NI FAs remains unclear. We hope these results provide guidance for the appropriate use and selection of reanalysis products in TC research and motivate further improvements in reanalysis systems.

## Conflict of Interest

The authors declare no conflicts of interest relevant to this study.

## Acknowledgments

Funding for this work is provided by the PCMDI project at Lawrence Livermore National Laboratory (LLNL), which is funded by the U.S. Department of Energy Office of Science Earth and Environmental System Modeling Program, which is part of the Earth and Environmental Systems Sciences Division of the Office of Biological and Environmental Research. Work by PAU was performed under the auspices of the U.S. Department of Energy by the LLNL (contract no. DE-AC52-07NA27344). C.Z. and S.B. were supported by the U.S. National Science Foundation (grant AGS-2244919) and the U.K. Natural Environment Research Council (NERC Large Grant NE/W009587/1), respectively, through the jointly funded Project Huracán.

## Open Research

The latest version (version 2.4.2) of TempestExtremes (TE) can be accessed from https://doi.org/10.5281/zenodo.4016883 (Ullrich et al., 2026). The latest version (version 1.2.1) of SyCLoPS can be assessed from https://doi.org/10.5281/zenodo.18287268 (Han, 2026). ERA5 TC track data can be found on https://doi.org/10.5281/zenodo.10906284 (Han & Ullrich, 2026a). JRA-3Q and JRA-55 TC track data can be accessed from https://figshare.com/s/bbb31ad78e6d51902505. The ERA5 dataset was obtained from the Research Data Archive (European Centre for Medium-Range Weather Forecasts, 2019) at the National Center for Atmospheric Research (https://doi.org/10.5065/BH6N-5N20). The IBTrACS archive (Gahtan et al., 2024) can be retrieved from https://www.ncei.noaa.gov/products/international-best-track-archive.

## Supporting Information

**Text S1**. Introduction of Hybrid TC Label in SyCLoPS

Many of the missed North Atlantic TC detections in ERA5 are classified as subtropical systems because they fail to satisfy the RH100MAX criterion (maximum relative humidity at 100 hPa within 2.0°) in SyCLoPS's Tropical Condition. Since the thresholds defining the tropical condition were trained using the global TC population, the RH100MAX threshold may be overly restrictive for the North Atlantic, leading to a very low FAR but also a comparatively low HR. To evaluate this possibility, we perform a sensitivity analysis by lowering the original RH100MAX threshold (RH100MAX > 20%) in 5% increments while keeping all other TC classification criteria unchanged. The results indicate that both ERA5 and JRA-3Q achieve their best overall performance with a relaxed threshold of RH100MAX > 5%. Under this criterion, the HR for North Atlantic TCs increases in both reanalyses, with only a comparatively modest increase in FAR. In contrast, JRA-55 exhibits only marginal improvements in HR across all tested thresholds while incurring a substantially larger increase in FAR. Based on these results, SyCLoPS assigns the "Hybrid TC" label to low-pressure systems that satisfy the TC classification criteria under this modified tropical condition.

Figure S2 shows the spatial distribution of global hybrid TCs in hybrid TC tracks, defined as tracks containing at least four 6-hourly Hybrid TC nodes. In both ERA5 and JRA-3Q, Hybrid TCs occur most frequently over the subtropical North Atlantic, corresponding closely to many observed TCs that were previously missed by the standard detection algorithm. Because hybrid TC tracks are rare outside the North Atlantic, including these systems produces little change in the HR and FAR of either reanalysis in the other basins. On the other hand, hybrid TCs in JRA-55 are much more spatially dispersed, likely explaining its poor performance when hybrid TCs are included in the North Atlantic TC evaluation. The discrepancy among reanalyses reflects the inherent difficulty of objectively identifying hybrid TCs, whose characteristics lie near the classification boundaries of both operational agencies and objective tracking algorithms.

These results suggest that future users of SyCLoPS should carefully consider whether to include hybrid TCs when evaluating North Atlantic TCs in other datasets. Before doing so, users should verify that the dataset reproduces a realistic spatial distribution of hybrid TCs comparable to that found in ERA5 and JRA-3Q. Otherwise, including hybrid TCs may lead to worse TC detection performance.

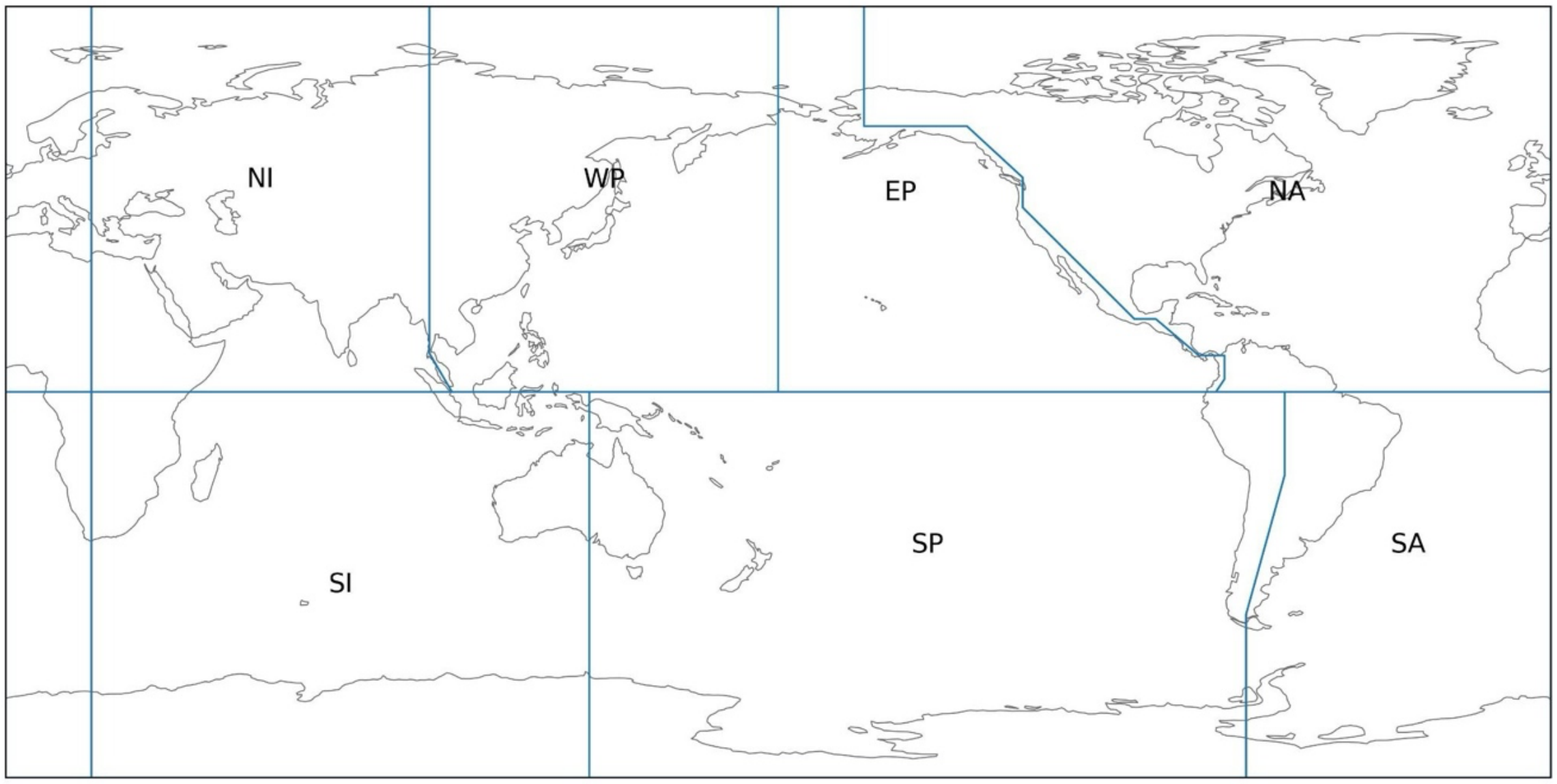

**Figure S1**. Basin boundaries used in this study.

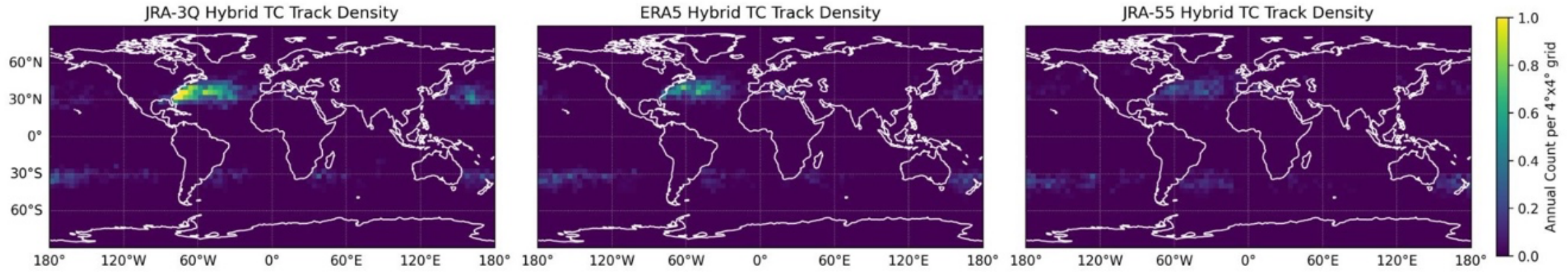

**Figure S2**. Hybrid TC 4° × 4° 1958-2023 annual-mean track density.

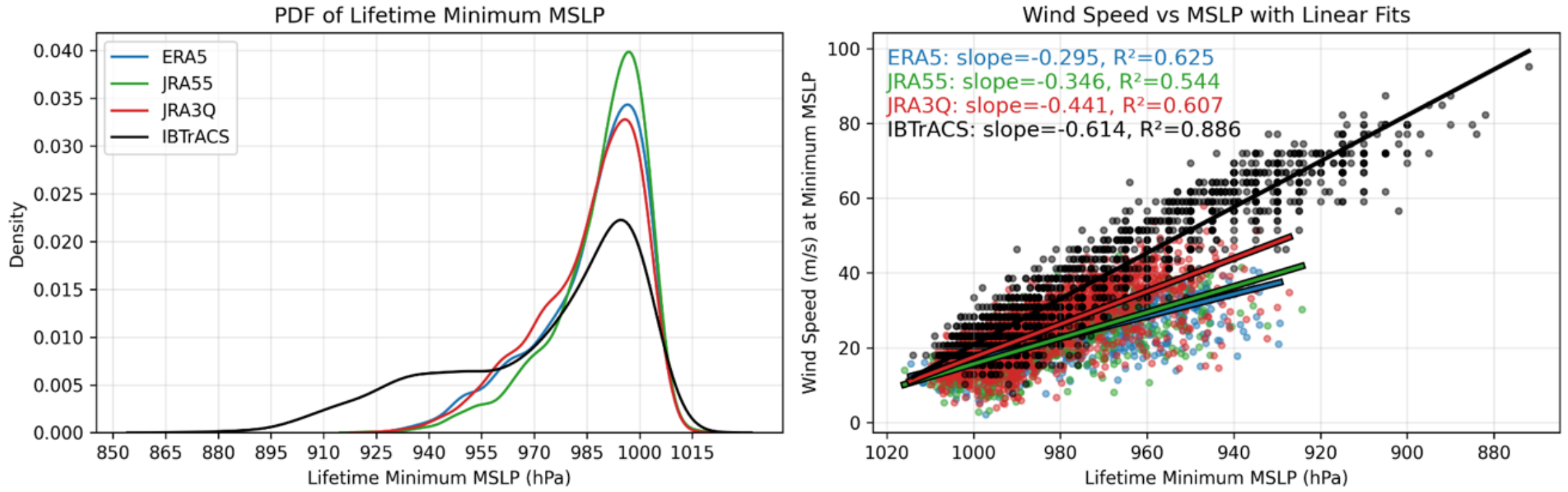

**Figure S3**. TC intensity representation in ERA5 (blue), JRA-55 (green), and JRA-3Q (red) since 2000.

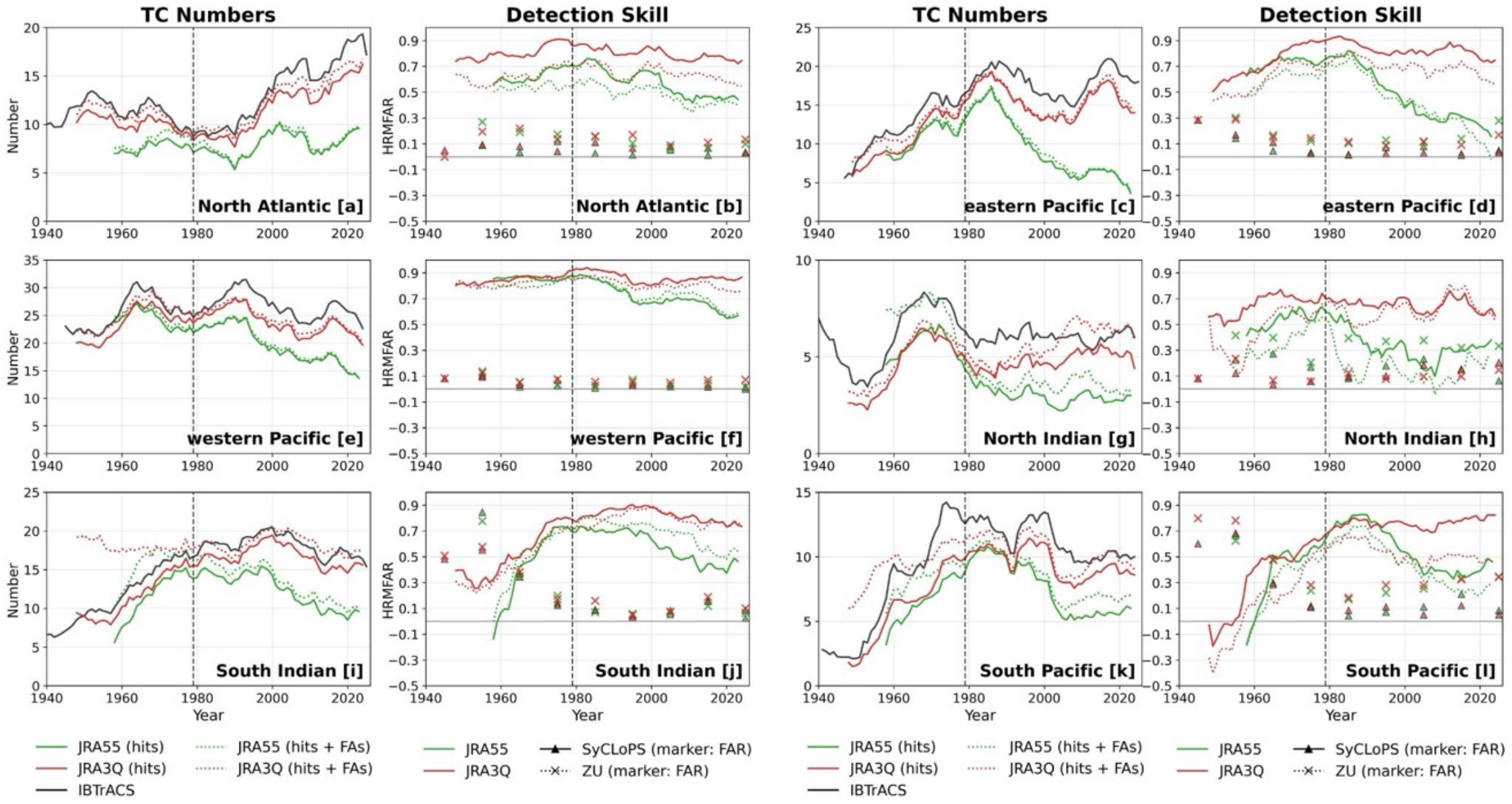

**Figure S4**. As in Figure 2, but for comparison between JRA-55 and JRA-3Q.

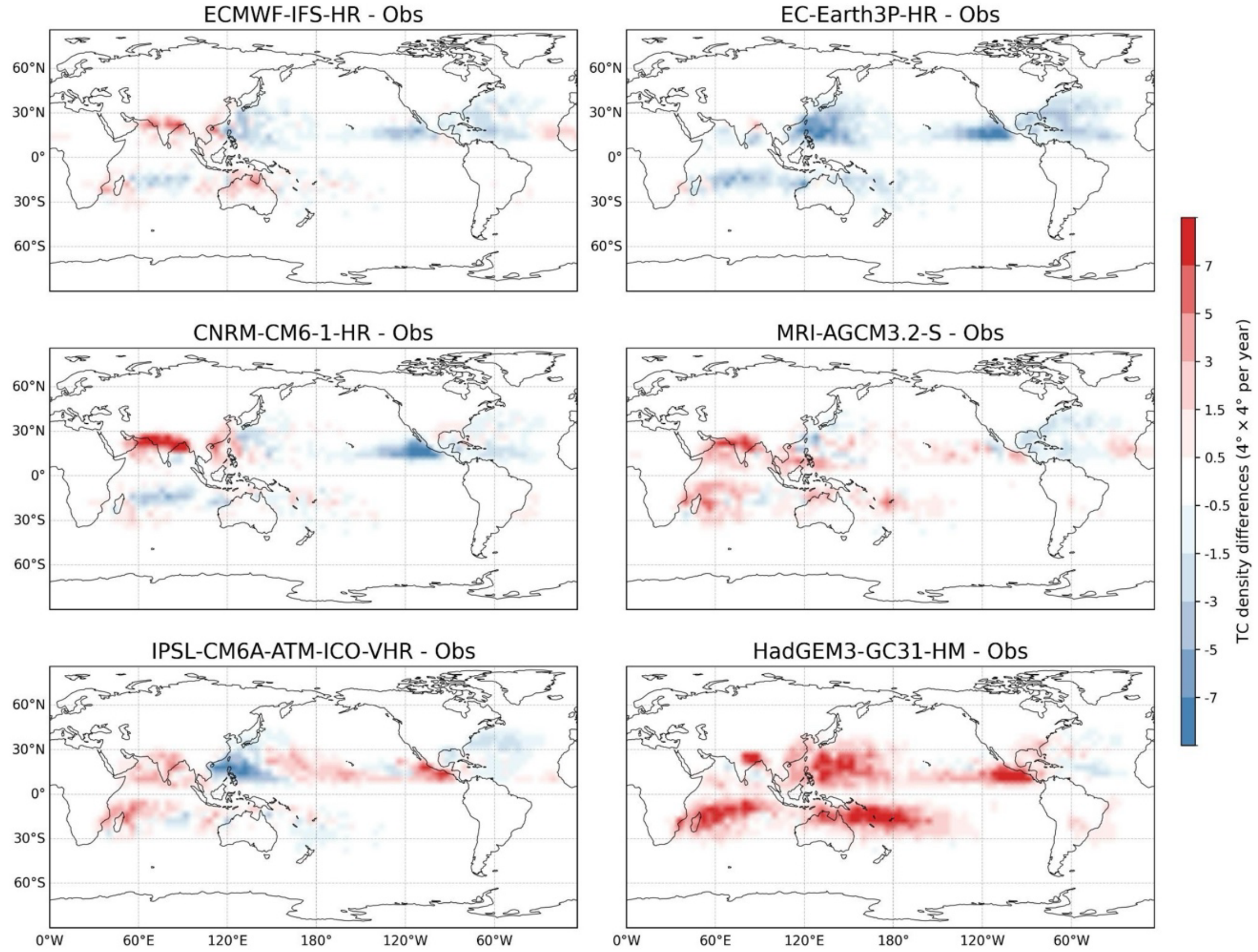

**Figure S5**. TC 4° × 4° annual-mean density differences between HighResMIP models and IBrACS over 1988-2014.

| | ERA5 | JRA-3Q | JRA-55 |
|---|---|---|---|
| USA | HR: 78%; FAR: 11% | HR: 90%; FAR: 6% | HR: 65%; FAR: 5% |
| WMO | HR: 80%; FAR: 14% | HR: 91%; FAR: 10% | HR: 67%; FAR: 8% |
| WMO+USA | HR: 77%; FAR: 11% | HR: 90%; FAR: 6% | HR: 65%; FAR: 5% |

**Table S1.** TC detection skill achieved by SyCLoPS in each reanalysis product based on different IBTrACS agency sources. “USA” denotes the source used in this study; “WMO” denotes data from the designated WMO agency in each basin; and “WMO+USA” primarily uses WMO agency data, supplemented by USA agency data when WMO data are unavailable. The analysis period is 1979–2024 for ERA5 and JRA-3Q and 1979–2023 for JRA-55. The slightly high FAR in “WMO” may be explained by the fact that some regional meteorological agencies did not accept their official WMO responsibility until the early 1990s.

## References

Alet, F., Andersson, T. R., Price, I., Markou, S., El-Kadi, A., Masters, D., Li, A., Merchant, S., Williams, N., Thornton, G., MacKay, K., Graham, O., Uddin, A., Gaiarin, B., Shah, D., Kruse, E., Hogsett, W., Zelinsky, D., Cangialosi, J., … Battaglia, P. (2026). Operational Tropical Cyclone Forecasting with AI. *Nature*, 1–3. https://doi.org/10.1038/s41586-026-10953-2

Bell, B., Hersbach, H., Simmons, A., Berrisford, P., Dahlgren, P., Horányi, A., Muñoz-Sabater, J., Nicolas, J., Radu, R., Schepers, D., Soci, C., Villaume, S., Bidlot, J.-R., Haimberger, L., Woollen, J., Buontempo, C., & Thépaut, J.-N. (2021). The ERA5 global reanalysis: Preliminary extension to 1950. *Quarterly Journal of the Royal Meteorological Society*, *147*(741), 4186–4227. https://doi.org/10.1002/qj.4174

Bhatia, K., Baker, A., Yang, W., Vecchi, G., Knutson, T., Murakami, H., Kossin, J., Hodges, K., Dixon, K., Bronselaer, B., & Whitlock, C. (2022). A potential explanation for the global increase in tropical cyclone rapid intensification. *Nature Communications*, *13*(1), 6626. https://doi.org/10.1038/s41467-022-34321-6

Bian, G.-F., Nie, G.-Z., & Qiu, X. (2021). How well is outer tropical cyclone size represented in the ERA5 reanalysis dataset? *Atmospheric Research*, *249*, 105339. https://doi.org/10.1016/j.atmosres.2020.105339

Bourdin, S. (2026). *RABTracks V1* (Version V1) [Dataset]. Zenodo. https://doi.org/10.5281/ZENODO.20426510

Bourdin, S., Fromang, S., Dulac, W., Cattiaux, J., & Chauvin, F. (2022). Intercomparison of four algorithms for detecting tropical cyclones using ERA5. *Geoscientific Model Development*, *15*(17), 6759–6786. https://doi.org/10.5194/gmd-15-6759-2022

Carstens, J. D., Didlake, A. C., & Zarzycki, C. M. (2024). Tropical Cyclone Wind Shear-Relative Asymmetry in Reanalyses. *Journal of Climate*, *37*(22), 5793–5861. https://doi.org/10.1175/JCLI-D-23-0628.1

Carstens, J. D., & Wing, A. A. (2022). Simulating Dropsondes to Assess Moist Static Energy Variability in Tropical Cyclones. *Geophysical Research Letters*, *49*(15), e2022GL099101. https://doi.org/10.1029/2022GL099101

Chand, S. S., Walsh, K. J. E., Camargo, S. J., Kossin, J. P., Tory, K. J., Wehner, M. F., Chan, J. C. L., Klotzbach, P. J., Dowdy, A. J., Bell, S. S., Ramsay, H. A., & Murakami, H. (2022). Declining tropical cyclone frequency under global warming. *Nature Climate Change*, *12*(7), 655–661. https://doi.org/10.1038/s41558-022-01388-4

Chavas, D. R., Camargo, S. J., & Tippett, M. K. (2025). Tropical Cyclone Genesis Potential Using a Ventilated Potential Intensity. *Journal of Climate*, *38*(7), 1667–1689. https://doi.org/10.1175/JCLI-D-24-0186.1

CRED and UNDRR. (2020). Human cost of disasters : An overview of the last 20 years (2000-2019) [Report]. https://www.undrr.org/publication/ human-cost-disasters-overview-last-20-years-2000-2019

Dirkes, C. A., Wing, A. A., Camargo, S. J., & Kim, D. (2023). Process-Oriented Diagnosis of Tropical Cyclones in Reanalyses Using a Moist Static Energy Variance Budget. *Journal of Climate*, *36*(16). https://doi.org/10.1175/JCLI-D-22-0384.1

Dulac, W., Cattiaux, J., Chauvin, F., Bourdin, S., & Fromang, S. (2024). Assessing the representation of tropical cyclones in ERA5 with the CNRM tracker. *Climate Dynamics*, *62*(1), 223–238. https://doi.org/10.1007/s00382-023-06902-8

European Centre for Medium-Range Weather Forecasts. (2019). ERA5 Reanalysis (0.25 Degree Latitude-Longitude Grid) [Dataset]. *NSF National Center for Atmospheric Research*. https://doi.org/10.5065/BH6N-5N20

Gahtan, J., Knapp, K. R., Schrek, C. J., Diamond, H. J., & Kossin, J. P. (2024). *International Best Track Archive for Climate Stewardship (IBTrACS) Project, Version 4* [Dataset]. NOAA National Centers for Environmental Information. https://doi.org/10.25921/82TY-9E16

Han, Y. (2026). *yepkids/SyCLoPS: SyCLoPS v1.1.2* [Computer software]. Zenodo. https://doi.org/10.5281/zenodo.18287269

Han, Y., & Ullrich, P. (2026a). *The System for Classification of Low-Pressure Systems (SyCLoPS) Dataset (Based on ERA5)* (Version 9) [Dataset]. Zenodo. https://doi.org/10.5281/zenodo.21834162

Han, Y., & Ullrich, P. A. (2026b). The Evolution of Tropical Cyclone Intensity and Size in Observations and HighResMIP Models. *Journal of Geophysical Research: Atmospheres*, *131*(16), e2026JD046560. https://doi.org/10.1029/2026JD046560

Hersbach, H., Bell, B., Berrisford, P., Hirahara, S., Horányi, A., Muñoz-Sabater, J., Nicolas, J., Peubey, C., Radu, R., Schepers, D., Simmons, A., Soci, C., Abdalla, S., Abellan, X., Balsamo, G., Bechtold, P., Biavati, G., Bidlot, J., Bonavita, M., … Thépaut, J.-N. (2020). The ERA5 global reanalysis. *Quarterly Journal of the Royal Meteorological Society*, *146*(730), 1999–2049. https://doi.org/10.1002/qj.3803

Hodges, K., Cobb, A., & Vidale, P. L. (2017). How Well Are Tropical Cyclones Represented in Reanalysis Datasets? *Journal of Climate*, *30*(14), 5243–5264. https://doi.org/10.1175/JCLI-D-16-0557.1

Hunt, K. M. R., Turner, A. G., Inness, P. M., Parker, D. E., & Levine, R. C. (2016). On the Structure and Dynamics of Indian Monsoon Depressions. *Monthly Weather Review*, *144*(9), 3391–3416. https://doi.org/10.1175/MWR-D-15-0138.1

Knapp, K. R., Kruk, M. C., Levinson, D. H., Diamond, H. J., & Neumann, C. J. (2010). The International Best Track Archive for Climate Stewardship (IBTrACS): Unifying Tropical Cyclone Data. *Bulletin of the American Meteorological Society*, *91*(3), 363–376. https://doi.org/10.1175/2009BAMS2755.1

Kobayashi, S., Ota, Y., Harada, Y., Ebita, A., Moriya, M., Onoda, H., Onogi, K., Kamahori, H., Kobayashi, C., Endo, H., Miyaoka, K., & Takahashi, K. (2015). The JRA-55 Reanalysis: General Specifications and Basic Characteristics. *気象集誌. 第2輯*, *93*(1), 5–48. https://doi.org/10.2151/jmsj.2015-001

Kochkov, D., Yuval, J., Langmore, I., Norgaard, P., Smith, J., Mooers, G., Klöwer, M., Lottes, J., Rasp, S., Düben, P., Hatfield, S., Battaglia, P., Sanchez-Gonzalez, A., Willson, M., Brenner, M. P., & Hoyer, S. (2024). Neural general circulation models for weather and climate. *Nature*, *632*(8027), 1060–1066. https://doi.org/10.1038/s41586-024-07744-y

Kosaka, Y., Kobayashi, S., Harada, Y., Kobayashi, C., Naoe, H., Yoshimoto, K., Harada, M., Goto, N., Chiba, J., Miyaoka, K., Sekiguchi, R., Deushi, M., Kamahori, H., Nakaegawa, T., Tanaka, T. Y., Tokuhiro, T., Sato, Y., Matsushita, Y., & Onogi, K. (2024). The JRA-3Q Reanalysis. *Journal of the Meteorological Society of Japan. Ser. II*, *102*(1), 49–109. https://doi.org/10.2151/jmsj.2024-004

Kossin, J. P., Olander, T. L., & Knapp, K. R. (2013). Trend Analysis with a New Global Record of Tropical Cyclone Intensity. *Journal of Climate*, *26*(24), 9960–9976. https://doi.org/10.1175/JCLI-D-13-00262.1

Malakar, P., Kesarkar, A. p., Bhate, J. n., Singh, V., & Deshamukhya, A. (2020). Comparison of Reanalysis Data Sets to Comprehend the Evolution of Tropical Cyclones Over North Indian Ocean. *Earth and Space Science*, *7*(2), e2019EA000978. https://doi.org/10.1029/2019EA000978

Murakami, H. (2014). Tropical cyclones in reanalysis data sets. *Geophysical Research Letters*, *41*(6), 2133–2141. https://doi.org/10.1002/2014GL059519

Roberts, C. D., Senan, R., Molteni, F., Boussetta, S., & Keeley, S. (2017). *ECMWF ECMWF-IFS-HR model output prepared for CMIP6 HighResMIP* [Dataset]. Earth System Grid Federation. https://doi.org/10.22033/ESGF/CMIP6.2461

Roberts, M. J., Camp, J., Seddon, J., Vidale, P. L., Hodges, K., Vanniere, B., Mecking, J., Haarsma, R., Bellucci, A., Scoccimarro, E., Caron, L.-P., Chauvin, F., Terray, L., Valcke, S., Moine, M.-P., Putrasahan, D., Roberts, C., Senan, R., Zarzycki, C., & Ullrich, P. (2020). Impact of Model Resolution on Tropical Cyclone Simulation Using the HighResMIP–PRIMAVERA Multimodel Ensemble. *Journal of Climate*, *33*(7), 2557–2583. https://doi.org/10.1175/JCLI-D-19-0639.1

Schenkel, B. A., & Hart, R. E. (2012). An Examination of Tropical Cyclone Position, Intensity, and Intensity Life Cycle within Atmospheric Reanalysis Datasets. *Journal of Climate*, *25*(10), 3453–3475. https://doi.org/10.1175/2011JCLI4208.1

Slocum, C. J., Razin, M. N., Knaff, J. A., & Stow, J. P. (2022). Does ERA5 Mark a New Era for Resolving the Tropical Cyclone Environment? *Journal of Climate*, *35*(21), 7147–7164. https://doi.org/10.1175/JCLI-D-22-0127.1

Sobel, A. H., Wing, A. A., Camargo, S. J., Patricola-DiRosario, C. M., Vecchi, G. A., Lee, C., & Tippett, M. K. (2021). Tropical Cyclone Frequency. *Earth's Future*, *9*(12), e2021EF002275. https://doi.org/10.1029/2021EF002275

Soci, C., Hersbach, H., Simmons, A., Poli, P., Bell, B., Berrisford, P., Horányi, A., Muñoz-Sabater, J., Nicolas, J., Radu, R., Schepers, D., Villaume, S., Haimberger, L., Woollen, J., Buontempo, C., & Thépaut, J.-N. (2024). The ERA5 global reanalysis from 1940 to 2022. *Quarterly Journal of the Royal Meteorological Society*, *150*(764), 4014–4048. https://doi.org/10.1002/qj.4803

Truchelut, R. E., Hart, R. E., & Luthman, B. (2013). *Global Identification of Previously Undetected Pre-Satellite-Era Tropical Cyclone Candidates in NOAA/CIRES Twentieth-Century Reanalysis Data*. *52*(10), 2243–2259. https://doi.org/10.1175/JAMC-D-12-0276.1

Ullrich, P. A., Zarzycki, C. M., McClenny, E. E., Pinheiro, M. C., Stansfield, A. M., & Reed, K. A. (2021). TempestExtremes v2.1: A community framework for feature detection, tracking, and analysis in large datasets. *Geoscientific Model Development*, *14*(8), 5023–5048. https://doi.org/10.5194/gmd-14-5023-2021

Ullrich, P., mariellep, Chen, H., Stachowicz, K., zarzycki, Han, Y., Huang, K., & Hoyer, S. (2026). *ClimateGlobalChange/tempestextremes: V2.4.2* [Computer software]. Zenodo. https://doi.org/10.5281/zenodo.18274602

Vecchi, G. A., & Knutson, T. R. (2011). *Estimating Annual Numbers of Atlantic Hurricanes Missing from the HURDAT Database (1878–1965) Using Ship Track Density*. https://doi.org/10.1175/2010JCLI3810.1

Vishnu, S., Boos, W. R., Ullrich, P. A., & O'Brien, T. A. (2020). Assessing Historical Variability of South Asian Monsoon Lows and Depressions With an Optimized Tracking Algorithm. *Journal of Geophysical Research: Atmospheres*, *125*(15), e2020JD032977. https://doi.org/10.1029/2020JD032977

Watt-Meyer, O., Henn, B., McGibbon, J., Clark, S. K., Kwa, A., Perkins, W. A., Wu, E., Harris, L., & Bretherton, C. S. (2025). ACE2: Accurately learning subseasonal to decadal atmospheric variability and forced responses. *Npj Climate and Atmospheric Science*, *8*(1), 205. https://doi.org/10.1038/s41612-025-01090-0

Xu, Z., Guo, J., Zhang, G., Ye, Y., Zhao, H., & Chen, H. (2024). Global tropical cyclone size and intensity reconstruction dataset for 1959–2022 based on IBTrACS and ERA5 data. *Earth System Science Data*, *16*(12), 5753–5766. https://doi.org/10.5194/essd-16-5753-2024

Ye, G., Leung, J. C.-H., Dong, W., Toumi, R., Xu, J., Li, W., Qian, W., Kong, H., & Zhang, B. (2026). A Reanalysis-Based Global Tropical Cyclone Tracks Dataset for the Twentieth Century (RGTracks-20C). *Earth System Science Data*, *18*(5), 3587–3607. https://doi.org/10.5194/essd-18-3587-2026

Zarzycki, C. M., & Ullrich, P. A. (2017). Assessing sensitivities in algorithmic detection of tropical cyclones in climate data. *Geophysical Research Letters*, *44*(2), 1141–1149. https://doi.org/10.1002/2016GL071606

Zarzycki, C. M., Ullrich, P. A., & Reed, K. A. (2021). Metrics for evaluating tropical cyclones in climate data. *Journal of Applied Meteorology and Climatology*. https://doi.org/10.1175/JAMC-D-20-0149.1

Zhang, K., & Chan, K. T. F. (2023). An ERA5 global climatology of tropical cyclone size asymmetry. *International Journal of Climatology*, *43*(2), 950–963. https://doi.org/10.1002/joc.7846

Zick, S. E., & Matyas, C. J. (2015). Tropical cyclones in the North American Regional Reanalysis: An assessment of spatial biases in location, intensity, and structure. *Journal* *of Geophysical Research: Atmospheres*, *120*(5), 1651–1669.
https://doi.org/10.1002/2014JD022417